\documentclass[conference]{IEEEtran}
\IEEEoverridecommandlockouts
\usepackage{geometry}
\usepackage{cite}
\usepackage{amsmath,amssymb,amsfonts}
\usepackage{algorithmic}
\usepackage[ruled,vlined]{algorithm2e}
\usepackage{graphicx}
\usepackage{textcomp}
\usepackage{xcolor}
\usepackage{bm}
\usepackage{eucal}
\usepackage{hyperref}
\usepackage{tikz}
\usepackage{stfloats}
\usetikzlibrary{arrows.meta, positioning, shapes.geometric, calc, fit, intersections, decorations.pathreplacing, decorations.pathmorphing}

\begin{document}
\title{Waveform Design for ISAC System: A Consensus ADMM Approach}
\author{%
  \IEEEauthorblockN{%
    Ngoc-Son Duong\IEEEauthorrefmark{1},
    Huyen-Trang Ta\IEEEauthorrefmark{1},
    Quang-Tang Ngo\IEEEauthorrefmark{1},
    Thi-Hue Duong\IEEEauthorrefmark{1}, 
    Van-Lap Nguyen\IEEEauthorrefmark{1},\\
    Cong-Minh Nguyen\IEEEauthorrefmark{2},
    Nguyen Minh Tran\IEEEauthorrefmark{1}, and
    Thai-Mai Dinh$^\#$\IEEEauthorrefmark{1}\thanks{This work has been done under the research project QG.25.06 “A Novel Algorithm for Joint Sensing and Communications in Next Generation Mobile Networks" of Vietnam National University, Hanoi.}
    \thanks{Ngoc-Son Duong is now with Faculty of Electronic Engineering, Posts and Telecommunications Institute of Technology, Hanoi, Vietnam, email: sondn@ptit.edu.vn}
  }
  \IEEEauthorblockA{\IEEEauthorrefmark{1}Faculty of Electronics and Telecommunications, VNU University of Engineering and Technology, Hanoi, Vietnam}%
  \IEEEauthorblockA{\IEEEauthorrefmark{2}National Foundation for Science and Technology Development, Hanoi, Vietnam}%
E-mails: $\{${sondn24, 22029064, 22029037, 22029040, 23020619, minhtran.nguyen, dttmai}$\rbrace$@vnu.edu.vn,\\ ncminh1805@gmail.com. ($^\#$\textit{Corresponding Author})}
\maketitle
\begin{abstract}
We study joint transmit-waveform and receive-filter design for a multi-user downlink integrated sensing and communication (ISAC) system under practical constant-modulus and similarity constraints. We cast the design as a unified multi-objective program that balances communication sum rate and sensing signal-to-interference-plus-noise ratio (SINR). To address this, we introduce an efficient algorithm that use consensus alternating direction method of multipliers (ADMM) framework to alternately update the transmit waveform and radar filter. The proposed method effectively handles the non-convex fractional sensing's SINR formulation and ensures fast convergence. Simulation results demonstrate that the proposed approach achieves better trade-offs between communication sum rate and sensing's SINR compared to existing benchmark schemes.
\end{abstract}
\section{Introduction}
The concept of integrated sensing and communication (ISAC) has emerged as a transformative approach for next-generation wireless systems, particularly in the context of 5G and beyond. ISAC systems enable the simultaneous use of the same waveform for both radar sensing and communication, offering significant advantages in terms of spectrum efficiency, latency reduction, and system resource sharing. By incorporating sensing and communication into a unified framework, ISAC holds the potential to significantly enhance the performance of next generation wireless systems, especially in applications like Internet of Things (IoT) and intelligence transportation systems.
However, optimizing performance in ISAC systems remains a big challenge due to the dual nature of the tasks involved. Both radar sensing and communication tasks have distinct requirements, such as the need for high sensitivity in radar target detection and robust signal reception for communication systems. Thus, designing a transmit waveform that meets both these objectives while managing interference and resource constraints is nontrivial. One of the primary challenges in ISAC design is waveform optimization, where the transmitted waveform must satisfy both radar sensing and communication objectives. In radar systems, this involves optimizing the waveform for target detection, while in communication systems, the focus shifts to efficient data transmission. Striking the right balance between these competing goals requires advanced techniques for managing the shared resources. 

Recent advancements in ISAC have been demonstrated across multiple studies. In particular, \cite{b1} introduces a penalty-based iterative beamformer optimization using block coordinate descent and weighted minimum mean square error for a full-duplex monostatic ISAC system, achieving up to 60 dB self-interference cancellation and significant improvements in both radar and communication performance. Following this direction, two cross-domain waveform optimization strategies—communication-centric and sensing-centric—are presented in \cite{b3}, jointly optimizing time, frequency, power, and delay-Doppler domains to suppress sidelobes, reduce peak-to-average power ratio (PAPR), and enhance both sensing accuracy and communication efficiency. Meanwhile, a hardware-efficient massive multiple-input multiple-output (MIMO) ISAC framework is developed in \cite{b2}, employing quantized constant-envelope constraints and low-resolution digital-to-analog converters, where an inexact augmented Lagrangian method with block successive upper-bound minimization effectively reduces beampattern mean squared error (MSE) and symbol error rate, highlighting the potential of massive MIMO for future radar performance enhancement.
In addition, \cite{b4} proposes two waveform designs—DSSS and OFDM—where DSSS with pseudo-random coding offers simplicity but suffers from Doppler and low data rates, whereas OFDM symbol-domain processing mitigates interference, accurately estimates multi-target range and velocity, and supports high-speed communication. To further enhance ISAC performance, \cite{b5} explores the integration of reconfigurable intelligent surface (RIS), jointly optimizing beamforming and RIS phase configuration to improve both target illumination power and user SINR, thereby maintaining effective operation even when the direct path is degraded or blocked. Finally, when the direct transmission path is obstructed, \cite{b6} demonstrates that coordinated optimization of beamforming and RIS phase for both radar and communication continues to enhance user signal-to-interference-plus-noise ratio (SINR) and radar target illumination, ensuring reliable dual-function performance even without a direct link.


In this paper, we present a novel waveform design framework for ISAC systems that jointly optimizes radar sensing and communication performance under practical hardware constraints using consensus alternating direction method of multipliers (ADMM) \cite{b11}. The designed waveform problem is formulated as a multi-objective optimization task that balances target detection capability and communication efficiency. To address its non-convex nature, a novel algorithm based on consensus ADMM is developed, enabling alternating optimization of the transmit waveform and receive filter while enforcing constant-modulus (CM) and similarity constraints. Simulation results verify that the proposed method achieves better trade-offs between radar and communication performance, demonstrating its practicality, scalability, and effectiveness for next-generation ISAC implementations.
\section{System model}
\begin{figure}
    \centering
    \includegraphics[width=1\linewidth]{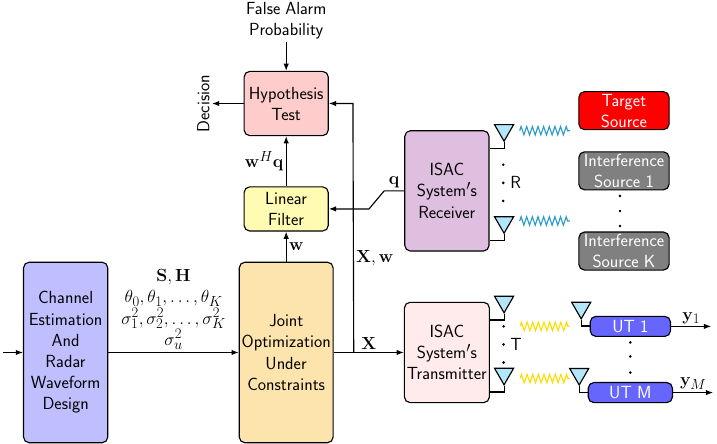}
    \caption{The ISAC system model with operation flow-chart.}
    \label{fig:sys_model}
\end{figure}
Fig. \ref{fig:sys_model} illustrates the overall architecture of a ISAC system in which the transmitter is equipped with a uniform linear array (ULA) of $T$ elements. The system simultaneously supports downlink communication to $M$ single-antenna user equipments (UE) and transmits radar probing waveforms to detect point-like targets. In addition, it is assumed that the ISAC system includes a dedicated receive ULA with $R$ elements, following the model of a radar system with collocated transmit and receive antenna arrays. We will provide a detailed presentation of the communication and radar functionalities in the following subsections.
\subsection{Communication model}
The signal matrix received at the communication user over the $N$ symbol intervals is expressed as
\begin{equation}
    \mathbf{Y} = \mathbf{HX} + \mathbf{Z},
\end{equation}
where $\mathbf{Y} \in \mathbb{C}^{M \times N}$ is the received signal matrix, and $y_{m,n}$ denotes the signal received at the $m$-th UE during the $n$-th symbol period. The transmit matrix $\mathbf{X} = [\mathbf{x}_1, \ldots, \mathbf{x}_N] \in \CMcal{X} \subset \mathbb{C}^{T \times N}$ consists of transmit vectors $\mathbf{x}_n \in \mathbb{C}^{T \times 1}$, where $\CMcal{X}$ defines the feasible set with spatial and temporal constraints. The channel matrix $\mathbf{H} = [\mathbf{h}_1, \ldots, \mathbf{h}_M]^{T} \in \mathbb{C}^{M \times T}$ models a flat-fading MIMO link, with each Gaussian random vector $\mathbf{h}_m \sim \CMcal{CN}(0, \mathbf{I}_T)$. The noise matrix $\mathbf{Z} = [\mathbf{z}_1, \ldots, \mathbf{z}_N] \in \mathbb{C}^{M \times N}$ represents additive white Gaussian noise (AWGN) at the receivers, where noise vector $\mathbf{z}_n \sim \CMcal{CN}(0, \sigma_z^2 \mathbf{I}_M)$ and $\sigma_z^2$ denotes the noise power. Let $\mathbf{s}_{m,n} \in \CMcal{O}$ denote the data symbol, which are drawn from a finite constellation $\CMcal{O}$. The desired symbol vector is represented as $\mathbf{s}_n = [s_{1,n},...,s_{M,n}]^\top \in \CMcal{O}^{M \times 1}$, and the entire symbol matrix is denoted as $\mathbf{S} = [\mathbf{s}_1,...,\mathbf{s}_N] \in \mathbb{C}^{M \times N}$. The communication objective is to design transmit vectors $\mathbf{x}_n$ such that the received signals $\mathbf{y}_n = \mathbf{H}\mathbf{x}_n + \mathbf{z}_n$ approximate the desired symbol vectors $\mathbf{s}_n$ as closely as possible. This can be formulated as the minimization of the total Multi-user Interference (MUI) energy
\begin{equation}
    f(\mathbf{X}) = \|\mathbf{HX} - \mathbf{S}\|^2_F,
\end{equation}
where $ \| \cdot \|_F$ denotes the Frobenius norm. 
\subsection{Radar model}
Let us consider the transmit signal vector $\mathbf{x}_n \in \mathbb{C}^{T \times 1}$ at the $n$-th symbol time, where $1 \leq n \leq N$, emitted by the ISAC system. The signal received at a location with azimuth angle $\theta$ is given by $\mathbf{a}^\top_t(\theta)\mathbf{x}_n, x \in N,$ where $\mathbf{a}_t(\theta) \in \mathbb{C}^{T \times 1}$ denotes the transmit steering vector. Under the ULA configuration, $\mathbf{a}_t(\theta)$ takes the form
\begin{equation}
    \mathbf{a}_t(\theta) = \frac{1}{\sqrt{T}}  [1,e^{-j\pi\sin(\theta)},\dots,e^{-j\pi(T-1)\sin(\theta)}]^\top.
\end{equation}
Assume that a target is located at angle $\theta_0$ and there are $K$ interference sources located at angles $\theta_k, k \in [K]$. The received baseband signal vector $\mathbf{q}_n \in \mathbb{C}^{R \times 1}$ at the $n$-th symbol time is given by
\begin{equation}
    \mathbf{q}_n = \alpha_0\mathbf{a}_r(\theta_0)\mathbf{a}^\top_t(\theta_0)\mathbf{x}_n + \sum^K_{k=1} \alpha_k\mathbf{a}_r(\theta_k)\mathbf{a}^\top_t(\theta_k)\mathbf{x}_n + \mathbf{u}_n,
\end{equation}
where $\alpha_0, \alpha_k$ are complex gains of the target ($\mathbb{E}\lbrace|\alpha_0|^2 \rbrace = \sigma_0^2$) and the $k$-th interference source ($\mathbb{E}\lbrace|\alpha_k|^2 \rbrace = \sigma_k^2$), respectively. The noise vector $\mathbf{u}_n \sim \CMcal{CN}(0, \sigma^2_u\mathbf{I}_R)$ is modeled as spatially white circular Gaussian noise.
The receive steering vector $\mathbf{a}_r(\theta) \in \mathbb{C}^{R \times 1}$ is defined analogously as
\begin{equation}
    \mathbf{a}_r(\theta) = \frac{1}{\sqrt{R}}[1, e^{-j\pi\sin(\theta)},\dots,e^{-j\pi(R - 1)\sin(\theta)}]^\top.
\end{equation}
Assuming that the angle of departure and arrival are identical (i.e., transmit and receive arrays are collocated), the received signal $\mathbf{q}$ can be rewritten as
\begin{equation}
    \mathbf{q} = \mathbf{A}(\theta_0)\mathbf{x} + \sum^K_{k = 1}\mathbf{A}(\theta_k)\mathbf{x} + \mathbf{u},
\end{equation}
where
\begin{equation}
    \mathbf{A}(\theta) = \mathbf{I}_N \otimes [\mathbf{a}_r(\theta)\mathbf{a}^\top_t(\theta)].
\end{equation}
A finite impulse response filter $\mathbf{w} \in \mathbb{C}^{R \times 1}$ is applied to the received signal $\mathbf{q}$ to form
\begin{equation}
    c = \mathbf{w}^H\mathbf{q} =\alpha_0\mathbf{w}^H\mathbf{A}(\theta_0)\mathbf{x} + \sum^K_{k=1}\alpha_k\mathbf{w}^H\mathbf{A}(\theta_k)\mathbf{x} + \mathbf{w}^H\mathbf{u}.
\end{equation}
The output sensing's SINR is expressed as
\begin{equation}
    \text{SINR}(\mathbf{x}, \mathbf{w}) = \frac{\sigma^2_0|\mathbf{w}^H\mathbf{A}(\theta_0)\mathbf{x}|^2}{\sum^K_{k = 1}\sigma^2_k|\mathbf{w}^H\mathbf{A}(\theta_k)\mathbf{x}|^2 + \sigma^2_u\|\mathbf{w}\|^2} .
\end{equation}
Intuitively, the waveform $\mathbf{x}$ and the receive filter $\mathbf{w}$ can thus be jointly designed by maximizing $\text{SINR}(\mathbf{x}, \mathbf{w})$. This maximization ensures enhanced target detection probability, while placing deep nulls in the directions of known interference sources. The beamforming effect arises naturally from this joint design, eliminating the need for predefined look directions. When environmental awareness is available, such as prior information on interference angles and powers, the optimization can be further improved via cognitive radar techniques.
\section{Problem Formulation and Proposed Method} 
\subsection{Problem Formulation}
The formulation simultaneously aims to minimize communication distortion, captured by a distortion metric $f'(\mathbf{x)} = \Vert\tilde{\mathbf{H}}\mathbf{x} - \mathbf{s}\Vert_2^2$, where $\tilde{\mathbf{H}} = \mathbf{H} \otimes\mathbf{I}$, $\mathbf{x} = \text{vec}(\mathbf{X})$ and $\mathbf{s} = \text{vec}(\mathbf{S})$; to maximize radar sensing performance, quantified via the output SINR $\text{SINR}(\mathbf{x},\mathbf{w})$; and to promote structural similarity between the synthesized and reference radar waveform $\mathbf{x}_0$. The resulting optimization problem is formulated as
\begin{equation}
\begin{aligned}
\min_{\mathbf{x},\, \mathbf{w}} \quad 
& \rho f'(\mathbf{x}) 
+ (1 - \rho)\frac{1}{\text{SINR}(\mathbf{x}, \mathbf{w})} + (1 - \rho)\lambda\Vert\mathbf{x} - \mathbf{x}_0\Vert_2^2
\\
\text{s.t.} \quad 
& |x_n| = \sqrt{\frac{P_{\max}}{TN}}, \quad \forall n
\end{aligned}
\label{org_cost}
\end{equation}
where $P_{\max}$ denotes total transmission power and the parameter $\rho \in [0,1]$ controls the trade-off between communication and sensing objectives. Meanwhile, the trade-off is further modulated by a regularization parameter $\lambda > 0$, which controls the emphasis placed on waveform similarity relative to other objectives. This soft constraint provides flexibility and enables compatibility with multi-objective waveform design formulations.
\subsection{Proposed method}
As all the problems are nonconvex but have differentiable objective functions with respect to both $\mathbf{w}$ and $\mathbf{x}$, we adopt the alternating minimization framework. The steps are given by
\begin{align}
\mathbf{x}^{(k+1)} &= \arg\min_{\mathbf{x}} g(\mathbf{x}, \mathbf{w}^{(k)}),  \label{eq:x_update}\\
\mathbf{w}^{(k+1)} &= \arg\min_{\mathbf{w}}(1 - \rho) \frac{1}{\text{SINR}(\mathbf{x}^{(k+1)}, \mathbf{w})}, \label{eq:w_update}
\end{align}
where $\mathbf{x}^{(k)}$ and $\mathbf{w}^{(k)}$ are the solutions available to problems \eqref{eq:x_update} and \eqref{eq:w_update}, respectively, at the $k$-th iteration. Besides, $g(\mathbf{x}, \mathbf{w}^{(k)})$ is exact \eqref{org_cost} with known $\mathbf{w}$.
\subsection{Optimize $\mathbf{w}$ when knowing $\mathbf{x}$}
Let $|\mathbf{w}^H\mathbf{A}(\theta_0)\mathbf{x}| = 1$, the closed-form solution of $\mathbf{w}$ is given by \cite{b12}
\begin{equation}
    \mathbf{w} = \frac{\mathbf{B}^{-1}\mathbf{a}}{\mathbf{a}^H\mathbf{B}^{-1}\mathbf{a}},
\end{equation}
where $\mathbf{B} = \sum_{k=1}^K \sigma_k^2 \big(\mathbf{A}(\theta_k) \mathbf{x}\big)\big(\mathbf{A}(\theta_k) \mathbf{x}\big)^H + \sigma_u^2\mathbf{I}$ and $\mathbf{a} = \mathbf{A}(\theta_0)\mathbf{x}$.
\subsection{Optimize $\mathbf{x}$ when knowing $\mathbf{w}$}
Introduce (w.r.t.\ $\mathbf{w}$) $\mathbf{R}_t \triangleq \sigma_0^2\,\mathbf{A}(\theta_0)^H\mathbf{w}\mathbf{w}^H\mathbf{A}(\theta_0)$, $
\mathbf{R}_i \triangleq \sum_{k=1}^{K}\sigma_k^2\,\mathbf{A}(\theta_k)^H\mathbf{w}\mathbf{w}^H\mathbf{A}(\theta_k)$ so that $\text{SINR}(\mathbf{x},\mathbf{w})
=\dfrac{\mathbf{x}^H\mathbf{R}_t\mathbf{x}}{\mathbf{x}^H\mathbf{R}_i\mathbf{x}+\sigma_u^2\|\mathbf{w}\|_2^2}$. It is easy to see that minimizing $1/\gamma$ is equivalent to minimizing $\mathbf{x}^H\mathbf{R}_i\mathbf{x}$ subject to $\mathbf{x}^H\mathbf{R}_t\mathbf{x}=\sigma_0^2$; however, the CM constraint prevents a direct solution. We, therefore, split the objective into three blocks by introducing consensus copies $\mathbf{x}_c, \mathbf{x}_s, \mathbf{x}_b$ for the communication term, sensing term, and similarity term, respectively. The problem \eqref{org_cost} is re-written as
\begin{equation}
\begin{aligned}
\min \quad 
& \rho \Vert\tilde{\mathbf{H}}\mathbf{x}_c - \mathbf{s}\Vert_2^2 \\
& + (1 - \rho)\mathbf{x}_s^H \mathbf{R}_i\mathbf{x}_s + (1 - \rho)\lambda\Vert\mathbf{x}_b - \mathbf{x}_0\Vert_2^2
\\
\text{s.t.} \quad 
& \mathbf{x}_c = \mathbf{x}, \mathbf{x}_s = \mathbf{x}, \mathbf{x}_b = \mathbf{x} \\
& \mathbf{x}_s^{H}\mathbf{R}_t\mathbf{x}_s = \sigma_0^2 \\
& |x_n| = c, \quad \forall n \\
\end{aligned}
\label{consensus_cost}
\end{equation}
Ignore the last two constraints, the augmented Lagrangian corresponding is given by
\begin{equation}
\begin{split}
\CMcal{L}_\rho(\mathbf{x}_c,\mathbf{x}_s, \mathbf{x}_b,\bm{\mu},\mathbf{x}) = \\
\rho (\tilde{\mathbf{H}}\mathbf{x}_c - \mathbf{s}) + \bm{\mu}_c^H(\mathbf{x}_c-\mathbf{x}) + \frac{\gamma}{2}\,\|\mathbf{x}_c-\mathbf{x}\|_2^{2} \\
 + (1 - \rho)\mathbf{x}_s^H \mathbf{R}_i\mathbf{x}_s + \bm{\mu}_s^H(\mathbf{x}_s-\mathbf{x}) + \frac{\gamma}{2}\,\|\mathbf{x}_s-\mathbf{x}\|_2^{2} \\
+ (1 - \rho)\lambda\Vert\mathbf{x}_b - \mathbf{x}_0\Vert_2^2 + \bm{\mu}_b^H(\mathbf{x}_b-\mathbf{x}) + \frac{\gamma}{2}\,\|\mathbf{x}_b-\mathbf{x}\|_2^{2}
\end{split}
\end{equation}

The resulting iterations are given by
\begin{multline}
\mathbf{x}^{(t+1)}_c = \arg\min_{\mathbf{x}_c} \lbrace \rho (\tilde{\mathbf{H}}\mathbf{x}_c - \mathbf{s}) + {\bm{\mu}_c^H}^{(t)}\big(\mathbf{x}_c-\mathbf{x}^{(t)}\big) \\ + \frac{\gamma}{2}\,\|\mathbf{x}_c -\mathbf{x}^{(t)}\|_2^{2}\rbrace.
\label{xc}
\end{multline}

\begin{multline}
\mathbf{x}^{(t+1)}_s = \arg\min_{\mathbf{x}_s} \lbrace (1 - \rho)\mathbf{x}_s^H \mathbf{R}_i\mathbf{x}_s + {\bm{\mu}_s^H}^{(t)}\big(\mathbf{x}_s - \mathbf{x}^{(t)}\big) +\\ \frac{\gamma}{2}\,\|\mathbf{x}_s-\mathbf{x}^{(t)}\|_2^{2} \rbrace \quad \text{s.t.} \quad \mathbf{x}_s^{H}\mathbf{R}_t\mathbf{x}_s = \sigma_0^2
\label{xs}
\end{multline}

\begin{multline}
\mathbf{x}^{(t+1)}_b = \arg\min_{\mathbf{x}_b} \lbrace(1 - \rho)\lambda\Vert\mathbf{x}_b - \mathbf{x}_0\Vert_2^2  \\ + {\bm{\mu}_b^H}^{(t)}\big(\mathbf{x}_b -\mathbf{x}^{(t)}\big) + \frac{\gamma}{2}\,\|\mathbf{x}_b -\mathbf{x}^{(t)}\|_2^{2} \rbrace.
\label{xb}
\end{multline}

\begin{equation}
\mathbf{x}^{(t+1)} = \arg\min_{\mathbf{x}} \sum_{i=1}^{N}
\left\{ {\bm{\mu}_i^H}^{(t)}(-\mathbf{x}) + \frac{\gamma}{2}\Vert\mathbf{x}_i^{(t+1)} - \mathbf{x} \Vert_2^2\right\}.
\label{globe_update}
\end{equation}

\begin{equation}
\bm{\mu}_i^{(t+1)} = \bm{\mu}_i^{(t)} + \gamma\big(\mathbf{x}_i^{(t+1)}-\mathbf{x}^{(t+1)}\big), \quad i \in \lbrace c, s, b \rbrace
\end{equation}
\subsubsection{Solution for \eqref{xc}}
We, herein, use the scaled ADMM \cite{b14} to solve it. Let $\mathbf{u}_c^{(t)} = \frac{1}{\gamma}{\bm{\mu}}_c^{(t)}$, \eqref{xc} can be re-written as
\begin{equation}
\mathbf x_c^{(t+1)} = \arg\min_{\mathbf{x}_c} \Vert \tilde{\mathbf{H}} \mathbf{x}_c - \mathbf{s} \Vert_2^2 + \frac{\gamma}{2}\Vert \mathbf{x}_c - \mathbf{x}^{(t)} + \mathbf{u}_c^{(t)}\Vert_2^2.
\label{xc2}
\end{equation}
To find $\mathbf{x}_c$, we derive the derivative of right hand side (RHS) of \eqref{xc2} with respect to $\mathbf{x}_c$ and set it to 0
\begin{equation}
\tilde{\mathbf{H}}^H(\tilde{\mathbf{H}}\mathbf{x}_c - \mathbf{s}) + \frac{\gamma}{2}(\mathbf{x}_c - \mathbf{x}^{(t)} + \mathbf{u}_c^{(t)}) = 0.
\end{equation}
Then, we get
\begin{equation}
    \mathbf{x}_c^{(t+1)} = \left(\tilde{\mathbf{H}}^H\tilde{\mathbf{H}} + \frac{\gamma}{2}\mathbf{I}\right)^{-1} \left(\tilde{\mathbf{H}}^H\mathbf{s} + \frac{\gamma}{2} \mathbf{x}^{(t)} - \frac{\gamma}{2}\mathbf{u}_c^{(t)}\right).
    \label{udxc}
\end{equation}
\subsubsection{Solution for \eqref{xs}}
The equality constraint makes problem \eqref{xs} non-convex. Solving it directly is not easy and so we relax the equality constraint with an inequality, as follows
\begin{multline}
\mathbf{x}^{(t+1)}_s = \arg\min_{\mathbf{x}_s} \lbrace (1 - \rho)\mathbf{x}_s^H \mathbf{R}_i\mathbf{x}_s + {\bm{\mu}_s^H}^{(t)}\big(\mathbf{x}_s - \mathbf{x}^{(t)}\big) +\\ \frac{\gamma}{2}\,\|\mathbf{x}_s-\mathbf{x}^{(t)}\|_2^{2} \rbrace \quad \text{s.t.} \quad \mathbf{x}_s^{H}\mathbf{R}_t\mathbf{x}_s \leq \sigma_0^2.
\end{multline}
This problems is classifed as quadratically constrained quadratic program (QCQP). Let $\mathbf{u}_s^{(t)}=\frac{1}{\gamma}\bm\mu_s^{(t)}$, we have
\begin{multline}
\mathbf{x}^{(t+1)}_s = \arg\min_{\mathbf{x}_s} \lbrace (1 - \rho)\mathbf{x}_s^H \mathbf{R}_i\mathbf{x}_s + \frac{\gamma}{2}\,\|\mathbf{x}_s-\mathbf{x}^{(t)} +\mathbf{u}_s^{(t)}\|_2^{2} \rbrace \\ \quad \text{s.t.} \quad \mathbf{x}_s^{H}\mathbf{R}_t\mathbf{x}_s \leq \sigma_0^2
\end{multline}
The Lagrangian of the QCQP problem in the first place can be written as
\begin{equation}
\begin{aligned}
\CMcal{L}(\mathbf{x}_s, \tau)
= {} & (1-\rho)\,\mathbf{x}_s^H \mathbf{R}_i \mathbf{x}_s + \frac{\gamma}{2}\,
\bigl\|\mathbf{x}_s - \mathbf{x}^{(t)} + \mathbf{u}_s^{(t)}\bigr\|_2^{2} \\
& + \tau\,
\bigl(\mathbf{x}_s^H \mathbf{R}_t \mathbf{x}_s - \sigma_0^2\bigr).
\end{aligned}
\end{equation}
The KTT condition for this problem is
\begin{equation}
    \mathbf{R}_i\mathbf{x}_s + \frac{\gamma}{2}(\mathbf{x}_s-\mathbf{x}^{(t)} +\mathbf{u}_s^{(t)}) + \tau\mathbf{R}_t\mathbf{x}_s = 0,
\end{equation}
equivalently
\begin{equation}
    (\mathbf{R}_i + \frac{\gamma}{2}\mathbf{I} + \tau\mathbf{R}_t)\mathbf{x}_s = \frac{\gamma}{2}(\mathbf{x}^{(t)} -\mathbf{u}_s^{(t)})
\end{equation}
with primal feasibility $\mathbf{x}_s^H \mathbf{R}_t\mathbf{x}_s \leq \sigma_0^2$, dual feasibility $\tau \geq 0$ and complementary slackness $\tau(\mathbf{x}_s^H \mathbf{R}_t\mathbf{x}_s - \sigma_0^2) = 0$. Herein, we see that if $\tau = 0$, the solution is
\begin{equation}
    \mathbf{x}_s = (\mathbf{R}_i + \frac{\gamma}{2}\mathbf{I})^{-1}(\frac{\gamma}{2}(\mathbf{x}^{(t)} -\mathbf{u}_s^{(t)})).
    \label{eq30}
\end{equation}
We then check $\mathbf{x}_s$ for primal feasibility, if it is satisfied, we stop. Else, there exists a $\tau > 0$ such that
\begin{equation}
\begin{cases}
\mathbf{x}_s = (\mathbf{R}_i + \frac{\gamma}{2}\mathbf{I}+\tau\mathbf{R_t})^{-1}(\frac{\gamma}{2}(\mathbf{x}^{(t)} -\mathbf{u}_s^{(t)})) \\
\mathbf{x}_s(\tau)^{H} \mathbf{R}_t\mathbf{x}_s(\tau) = \sigma_0^2 
\end{cases}
\label{eq35}
\end{equation}
Let $q(\tau) = \mathbf{x}_s(\tau)^{H} \mathbf{R}_t\mathbf{x}_s(\tau)$, then $q(\tau)$ is strictly decreasing in $\tau$, hence the equation $q(\tau)=\sigma_0^2$ has a unique solution\ $\tau^\star>0$. In this case, one can use Bi-section search \cite{b16} to find $\tau^\star$.
\subsubsection{Solution for \eqref{xb}}
Let $\mathbf{u}_b^{(t)}=\frac{1}{\gamma}\bm\mu_b^{(t)}$, we have
\begin{equation}
\mathbf{x}_b^{(t+1)} =\arg\min_{\mathbf{x}_b}\ (1-\rho)\lambda \Vert\mathbf{x}_b-\mathbf{x}_0\Vert_2^2
+\frac{\gamma}{2}\Vert\mathbf{x}_b - \mathbf{x}^{(t)}+\mathbf{u}_b^{(t)}\Vert_2^2.
\label{xb2}
\end{equation}
To find $\mathbf{x}_b$, we derive the derivative of RHS of \eqref{xb2} with respect to $\mathbf{x}_b$ and set it to 0
\begin{equation}
    (1-\rho)\lambda(\mathbf{x}_b-\mathbf{x}_0) + \frac{\gamma}{2}(\mathbf{x}_b - \mathbf{x}^{(t)}+\mathbf{u}_b^{(t)})=0.
\end{equation}
Then, we get
\begin{equation}
\mathbf{x}_b^{(t+1)}=\frac{(1-\rho)\lambda\mathbf{x}_0 + \frac{\gamma}{2}\big(\mathbf{x}^{(t)} - \mathbf{u}_b^{(t)}\big)} {(1-\rho)\lambda+\frac{\gamma}{2}}
\label{udxb}
\end{equation}
\subsubsection{Solution for \eqref{globe_update}}
Let $m=3$ be the number of local blocks, the consensus step solves a least-squares average
\begin{equation}
\label{udx}
\mathbf{x}^{(t+1)}=\frac{1}{m}\sum_{i\in\{c,s,b\}}\big(\mathbf{x}_i^{(t+1)}+\mathbf{u}_i^{(t)}\big).
\end{equation}
Finally, we project $\mathbf{x}^{(t+1)}$ onto the CM set as \cite{b15}
\begin{equation}
x_n^{(t+1)} = \begin{cases}
\sqrt{\frac{P_{\max}}{TN}}, & x_n = 0\\
\sqrt{\frac{P_{\max}}{TN}} \frac{x_n^{(t+1)}}{|x_n^{(t+1)}|}, & x_n \neq 0
\end{cases}.
\label{prj}
\end{equation}
\begin{algorithm}[htbp]
\caption{Proposed consensus ADMM for finding $\mathbf{x}$ given $\mathbf{w}$}
\SetAlgoLined
\KwIn{$\tilde{\mathbf{H}}$, $\mathbf{s}$, $\mathbf{R}_i$, $\mathbf{R}_t$, $\mathbf{x}_0$, $\lambda$, $\rho \in [0,1)$, $\gamma>0$, ADMM tolerances $(\varepsilon_{\mathrm p}, \varepsilon_{\mathrm d})$}
\KwOut{$\mathbf{x}^\star$}

\textbf{Initialize:} $\mathbf{x}^{(0)}$; $\mathbf{u}_c^{(0)}=\mathbf{u}_s^{(0)}=\mathbf{u}_b^{(0)}=0$, $t\gets0$.

  \While{$1$}{
    \textit{\textbf{1) Local updates:}} \\
    - Update $\mathbf{x}_c^{(t+1)}$ via \eqref{udxc}\;
    - Update $\mathbf x_s^{(t+1)}$ via \eqref{eq30} or via solving \eqref{eq35}\;
    - Update $\mathbf{x}_b^{(t+1)}$ via \eqref{udxb}\;
    \textit{\textbf{2) Global consensus update:}} \\
    Update $\mathbf{x}^{(t+1)}$ via \eqref{udx}\;
    \textit{\textbf{3) Projection:}} \\
    - Update $x_n^{(t+1)}$ via \eqref{prj}\;
    \textit{\textbf{4) Dual update:}} \\
    $\mathbf{u}_i^{(t+1)} = \mathbf{u}_i^{(t)} + \big( \mathbf{x}_i^{(t+1)} - \mathbf{x}^{(t+1)} \big), \forall i \in \lbrace c, s, b\rbrace$\;
    
    \textit{\textbf{5) ADMM stopping check:}} \\
    $r^{(t+1)} \gets \Vert\mathbf{x}_c^{(t+1)} - \mathbf{x}^{(t+1)}\Vert + \Vert\mathbf{x}_s^{(t+1)} - \mathbf{x}^{(t+1)}\Vert + \Vert\mathbf{x}_b^{(t+1)} - \mathbf{x}^{(t+1)}\Vert$\;
    $s^{(t+1)} \gets \gamma\,\|\mathbf{x}^{(t+1)} - \mathbf{x}^{(t)}\|$\;
    \If{$(r^{(t+1)}\le \varepsilon_{\mathrm p})$ $\vee$ $(s^{(t+1)}\le \varepsilon_{\mathrm d})$}{
      \textbf{break}\;
    }
    $t\gets t+1$\;
  }
\end{algorithm}
\section{Simulation Results and Discussion}
\subsection{Simulation Setup}
In this section, numerical results are presented to evaluate the performance of the proposed technique. The ISAC transmitter is equipped with $N = 16$ antennas and serves $K = 4$ users over a communication frame of length $L = 20$, while aiming to detect a radar target located at the spatial angle $\theta_0 = 15^{\circ}$. Furthermore, two interference sources are assumed to be located at spatial angles $\theta_{1} = -50^{\circ}$ and $\theta_{2} = 40^{\circ}$. The transmit signal power is set to $P_T = 1$ W. The powers of the target and interference signals are set to $\sigma_0^2 = 10$ dB and $\sigma_k^2 = 30$ dB, respectively, while the noise variance at the radar receiver is $\sigma_u^2 = 0$ dB. The transmitter employs a ULA with half wavelength inter-element spacing. The transmitted symbols are generated from a quadrature phase-shift keying constellation with unit average power. The channel coefficients between the transmitter and the users are modeled as independent circularly symmetric complex Gaussian random variables following $\CMcal{CN}(0,1)$. The resulting performance is compared with several benchmark schemes, including orthogonal linear frequency modulation radar waveform, method in \cite{b15} and Zero-MUI.
\subsection{Simulation Results and Discussion}
\subsubsection{On the achievable sum-rate}
\begin{figure}[ht]
  \centering
  \includegraphics[width=0.95\columnwidth]{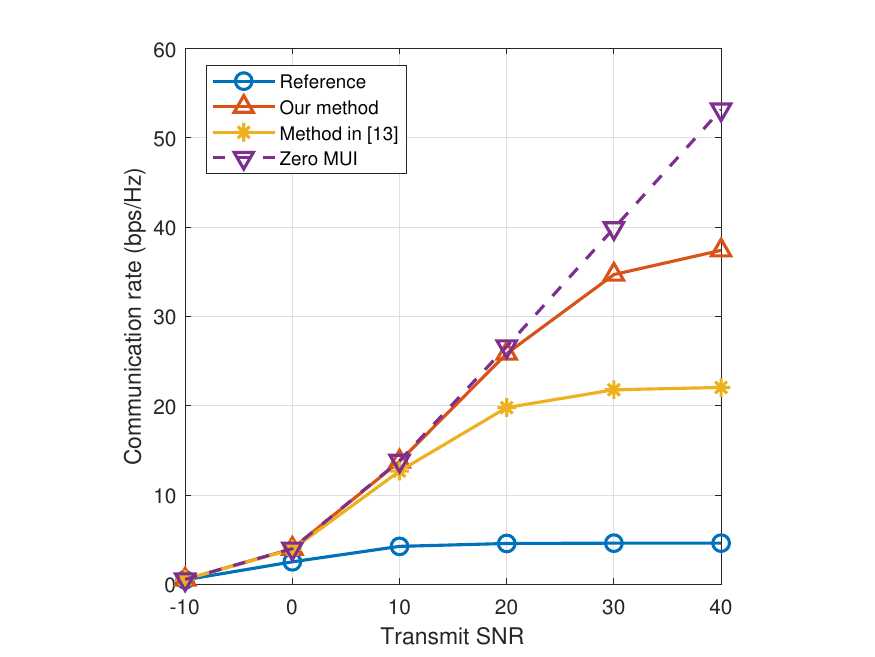}
  \caption{Sum-rate comparison for different approaches at $\rho = 0.2, \lambda = 1$.}
  \label{fig:sumrate}
\end{figure}

The sum-rate performance of three methods is illustrated by Fig. \ref{fig:sumrate}. As the transmit SNR increases, all curves show an upward trend, indicating improved communication efficiency at higher SNR levels. However, the proposed method achieves a consistently higher sum rate than both the radar waveform and method in \cite{b15} approaches over the entire SNR range. This demonstrates that the proposed joint waveform design effectively mitigates multi-user interference and allocates transmit power more efficiently. At high SNRs, the Zero MUI scheme attains the maximum rate, as it completely suppresses inter-user interference; nevertheless, this usually sacrifices radar functionality or increases implementation complexity. In contrast, our method achieves a performance level close to Zero MUI while maintaining practical system constraints, thus providing a desirable trade-off between communication throughput and radar compatibility. In additional, the method in \cite{b15} saturates at medium SNR values, reflecting its limited adaptability, whereas the radar waveform remains nearly flat, confirming that a non-optimized directional transmission cannot achieve high throughput. Overall, these results verify the superiority of the proposed design in enhancing communication performance within joint radar-communication systems.

\subsubsection{On the received sensing's SINR}
\begin{figure}[ht]
  \centering
  \includegraphics[width=0.9\columnwidth]{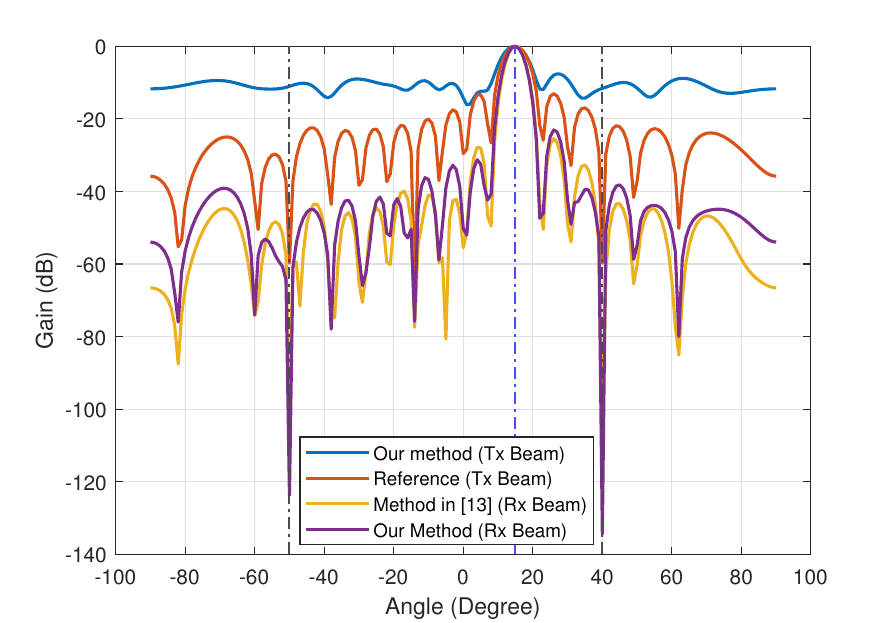}
  \caption{Beampatterns obtained by different approaches.}
  \label{fig:beampattern}
\end{figure}

\begin{figure}[ht]
  \centering
  \includegraphics[width=0.93\columnwidth]{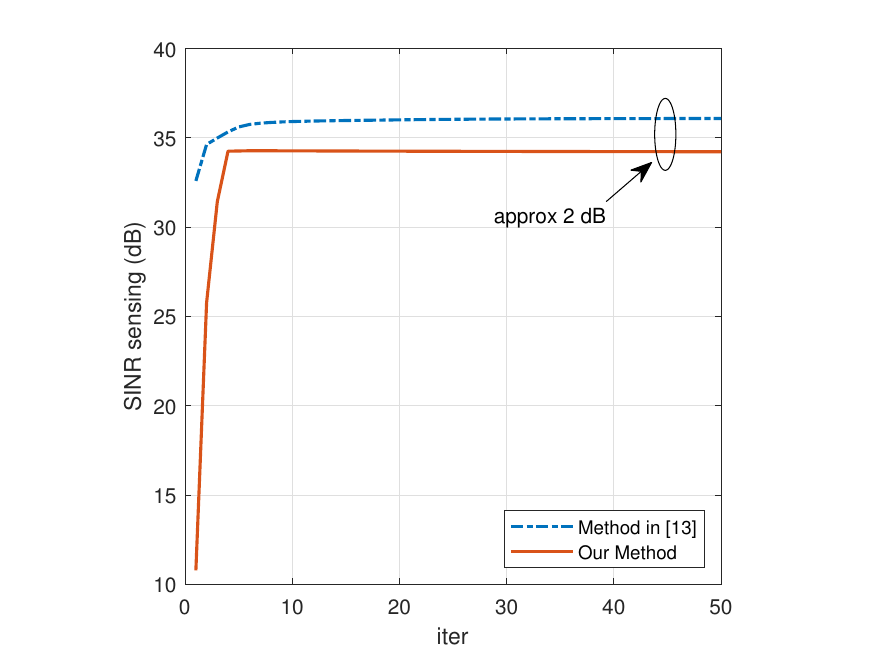}
  \caption{Sensing's SINR (dB) at $\rho = 0.2, \lambda = 1$.}
  \label{fig:SINR_r}
\end{figure}
The objective is to assess the capability of each approach to focus energy toward the desired target direction while suppressing interference in undesired directions, which is illustrated through the beampattern gain versus angle. The obtained results are depicted in Fig. \ref{fig:beampattern}, showing a detailed comparison of the received beampatterns for all methods. It can be observed that both methods maintain the main lobe toward the desired target direction with comparable peak gains, indicating that both systems effectively concentrate the transmitted energy on the target. However, some noticeable differences can still be observed. The proposed method exhibits interference nulls that are nearly identical in position and show significant attenuation, which helps improve the radar-received SINR and consequently enhances the target detection probability. In several angular regions, the side-lobe levels of the proposed method are slightly higher than method in \cite{b15}. This reflects a reasonable trade-off between beam focusing capability and communication performance (sum-rate). Overall, the proposed method achieves a good balance between energy concentration and interference suppression, demonstrating more flexible beam control that is suitable for ISAC designs with radar-prioritized performance. Fig. \ref{fig:SINR_r} shows the convergence of the proposed and method in \cite{b15}. Our method converges earlier in 10 iterations, indicating fast convergence. The proposed method has a steady-state SINR approximately 2 dB lower than \cite{b15} due to the rate–sensing constraint imposed.
\subsubsection{The overall trade-off}
\begin{figure}[htbp]
  \centering
  \includegraphics[scale=0.42,trim=60 220 10 210, clip]{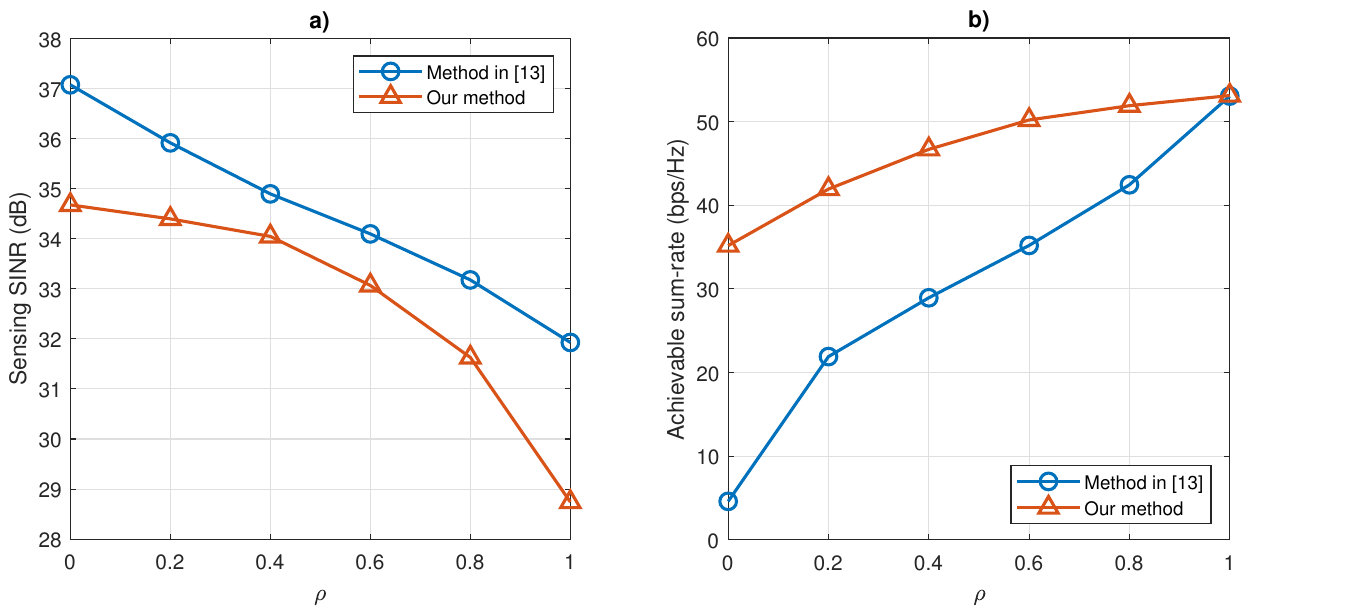}
  \caption{Sensing SINR and achievable sum-rate versus the trade-off factor $\rho \; (\lambda = 1)$. (a) Sensing SINR (dB). (b) Achievable sum-rate (bps/Hz).}
  \label{fig:tradeoff}
\end{figure}
Fig. \ref{fig:tradeoff} shows the trade-off between sensing and communication performance as  $\rho$ varies from $0$ to $1$ at $\lambda = 1$. Overall, both methods give relatively high sensing SINR for all $\rho$. The proposed method consistently achieves a higher sum-rate across all $\rho$ values while maintaining a smooth sensing SINR degradation ($1 - 3$ dB lower than \cite{b15}). This indicates robust and stable behavior under different trade-off settings. We observe that projected gradient (PG) \cite{b15} suffers from severe communication-rate degradation when $\rho$ is small (radar-dominated regime). This is because the combined gradient is dominated by the radar shaping term, so the waveform is optimized almost exclusively for target detection and interference suppression, while multi-user interference in the downlink is largely ignored. As a result, the sum rate collapses for $\rho \in [0,0.8]$. Only when $\rho = 1$, i.e., when the radar term is effectively deactivated, PG behaves like a conventional multi-user precoder design and achieves a high sum rate. In contrast, our consensus ADMM decouples the radar and communication objectives into separate subproblems and enforces their agreement via consensus, so the communication block remains influential even when $\rho$ is small. This yields significantly higher sum rates in the radar-dominated regime.
\section{Conclusion}
In this paper, we have proposed a novel consensus ADMM-based joint waveform and receive filter design framework for ISAC systems. The proposed method effectively addresses two critical practical constraints: the constant modulus requirement, which ensures compatibility with hardware limitations, and the similarity constraint, which preserves the desired radar beampattern structure. The novel algorithm based on consensus ADMM enables efficient handling of the fractional SINR expression and distributed convex subproblems, guaranteeing fast convergence. Simulation results confirm that the proposed approach achieves a better trade-off between radar and communication performance compared to benchmark methods. Future work will focus on the coupled challenges of designing a waveform that supports high uplink data rates while concurrently resolving the self-interference, thereby enabling a more complete system architecture.

\end{document}